\documentclass[11pt,aps,prd,preprint,preprintnumbers,showpacs,groupedaddress,floatfix]
{revtex4-1}
\usepackage{graphicx}
\usepackage{epsfig}
\usepackage{dcolumn}
\usepackage{bm}
\usepackage{subfigure}
\usepackage{amsmath}
\usepackage{amssymb}
\usepackage{bbm}
\usepackage{multirow}
\usepackage{setspace}
\usepackage{xcolor}
\usepackage[colorlinks,
            linkbordercolor={0 0 0},
            pdfborder={0 0 0},
            linkcolor=blue,
            citecolor=blue,
            urlcolor=blue,
            breaklinks=true]{hyperref}

\allowdisplaybreaks

\def\rm1 #1{\mbox{\scriptsize #1}}
\def\eq#1{\begin{equation} #1 \end{equation}}

\newlength{\x}
\newlength{\y}
\newlength{\z}
\urldef\milcurl\url{http://www.physics.utah.edu/~detar/milc/}

\begin{document}

\phantomsection
\addcontentsline{toc}{chapter}{Main}

\title{Meta-stable States in Quark-Gluon Plasma}
\author{Mridupawan Deka $^{a,b}$}
\email[e-mail:]{Mridupawan.Deka@physik.uni-regensburg.de}
\author{Sanatan Digal $^{a}$}
\email[e-mail:]{digal@imsc.res.in}
\author{Ananta P. Mishra $^{a}$}
\email[e-mail:]{apmishra@imsc.res.in}
\affiliation
{%
\centerline{$^{a}$ Institute of Mathematical Sciences, Chennai- 600113, India}\\
\centerline{$^{b}$ Institut f\"{u}r Theoretische Physik,\ Universit\"{a}t Regensburg,\ 
  D-93040 Regensburg,\ Germany}
}
\date{\today}

\bigskip

\begin{abstract}
\bigskip
We study the meta-stable states in high temperature phase of QCD characterised by 
nonzero expectation values for the imaginary part of the Polyakov loop.\ We consider 
$N_f= 2, 3$ dynamical staggered quarks,\ and carry out simulations at various values 
of the coupling $\beta$ to observe these states.\ In particular,\ we find the value of the 
coupling ($\beta_m$) above which the meta-stable states appear.\ The resulting value 
of $\beta_m$ corresponds to temperature $T_m \gtrsim 750$~MeV for
$N_f=2$.

\vspace{1pc}
\end{abstract}


\maketitle
\thispagestyle{empty}

\pagestyle{plain}
\setcounter{page}{1}
\pagenumbering{arabic}
\singlespacing
\parskip 5pt


\section{Introduction}

Experimental and theoretical studies of matter under extreme conditions is 
one of the active fields of research in recent times.\ Such a state of matter
is created in ultra-relativistic heavy-ion collisions.\ The fireball created in these 
collisions leads to a state of deconfined quarks and gluons (Quark-gluon plasma).\ 
With the increase in the collision energy,\ the fireball not only crosses the 
confinement-deconfinement transition temperature $T_c$ but probes deeper into 
the deconfined phase.\ For example,\ in the heavy-ion collisions at LHC the initial fireball 
temperature is expected to go up to $5 T_c$~\cite{Wiedemann:2009sa}.\ The increase 
in fireball temperature will result in the observation of new signals bearing the 
properties of the system at higher temperatures.\ Hence,\ it is important to study 
any possible prominent changes in the properties of the medium in the deconfinement 
phase away from $T_c$,\ which can be observed in experiments.\ In this context,\ we 
plan to study the explicit breaking of $Z(3)$ symmetry,\ and the meta-stable states 
associated with it.

In the pure $SU(3)$ gauge theory,\ the deconfined phase exists in three degenerate 
states characterised by three different values of the Polyakov loop.\ These
three states are related to each other via the $Z(3)$ rotations.\ So,\ in the
deconfined phase,\ $Z(3)$ symmetry is spontaneously broken.\ In the confinement
phase,\ the Polyakov loop average vanishes restoring the $Z(3)$ symmetry.\ 
For QCD with dynamical fermions,\ the $Z(3)$ symmetry is explicitly broken,\ and 
the degeneracy between the three states is lifted~\cite{Dixit:1991et}.\ Only the state for 
which the expectation value of Polyakov loop is real becomes the ground state.\ 
It is not clear what happens to the other two states.\ For 
asymptotically large temperatures,\ one expects the gluons to dominate so that the 
effects of quarks can at most make the other two states (with Polyakov loop
phase angle $\pm 2\pi/3$) meta-stable.

We would like to emphasize that the $Z(3)$ meta-stable states are not the results of 
meta-stability near any first order transition.\ They are different from the meta-stable 
states that one observes near a first order phase transition.\ A closer analogy of the 
$Z(3)$ meta-stable states would be a state of magnetization anti-parallel to the 
external field.

The $Z(3)$ meta-stable states are expected to play important role both in
the context of heavy-ion collision,\ and in the early Universe.\ If these states 
indeed exist just above $T_c$,\ then they can have significant effects on the medium 
properties~\cite{Wiedemann:2009sa}.\ If a fluctuation in the form of a meta-stable 
bubble in the back-ground of the stable phase costs free energy of the order of
the temperature scale~$T$,\ then such fluctuations will be present in the system.\ 
In the heavy-ion collision it is possible 
that the whole fireball may thermalise to one of these 
meta-stable states.\ This state will then decay through a first order phase 
transition even before the system cools down below $T_c$.\ It is also possible that the 
meta-stable phases of super-horizon size may occur in the early Universe,\ and 
decay through bubble nucleation~\cite{Ignatius:1991nk,Layek:2005zu}.

There are several studies on $Z(3)$ meta-stable states at high temperatures in the 
presence of quarks.\ It has been shown that the contribution of massless quarks to 
the one loop effective potential leads to meta-stable states for 
$T \geq T_c$~\cite{Dixit:1991et,Belyaev:1991cw}.\ 
These states have also been observed above the deconfinement 
transition in the Nambu-Jona-Lasinio model~\cite{Meisinger:1995ih}.\ There are only 
a very few lattice QCD studies on these meta-stable states.\ A lattice QCD study with 
fermions in the sextet representation has found meta-stable states, characterised 
by phase angle ($\pm 2\pi/3$, $\pi$), close to $T_c$ in the 
deconfinement phase~\cite{Machtey:2009wu}.\ However, in this work,\ we have 
considered $N_f=2, 3$ staggered fermions in the fundamental representation to look for 
the meta-stable states.\ Contrary to the previous studies,\ we find that the meta-stable 
states do not exist in the neighbourhood of $T_c$,\ but for 
temperatures~$T_m \gtrsim 750$~MeV.\ Though this temperature may not be reached at 
the SPS and RHIC experiments, there is a possibility that these states are accessible at 
LHC. The fact that $T_m > T_c$ may have important consequence(s) for the early Universe.

The paper is organized as follows.\ In section II,\ we review the $Z(3)$ 
symmetry and meta-stable states in the presence of dynamical fermions.\
Our lattice simulation techniques are discussed in section III.\ Results
are presented in section IV.\ We present our conclusions in 
section V.


\section{$Z(3)$ symmetry in the presence of dynamical quarks and meta-stable
states}

In this section,\ we briefly discuss the pure $SU(N)$ gauge theory at finite 
temperature,\ and their symmetries.\ Later,\ we focus on the case with dynamical 
quarks,\ and discuss their effects on the $Z(N)$ 
symmetry~\cite{'tHooft:1977hy,McLerran:1981pb}.

We start with the pure $SU(N)$ gauge theory at finite temperature,\
$T = \beta^{-1}_{\tau}$.\ In this case,\ one uses static fundamental charges (infinitely 
heavy test quarks) to probe into the dynamics of the pure glue system.\ The 
static fundamental charges are described by the Polyakov 
loop~\cite{Polyakov:1978vu,Susskind:1979up} which is defined as the trace of the 
thermal Wilson line,
\eq{
  \label{polyl}
  L(\vec x) = \frac{1}{N}\, {\mbox{Tr}}\, {\bf W}(\vec{\bf x}),
}
with the thermal Wilson line operator,\ ${\bf L}(\vec{\bf x})$,\ defined as,
\eq{
  \label{wilsonl}
  {\bf W}(\vec{\bf x}) = P\, \exp\left[i g \int^{\beta_{\tau}}_0 A_0(\vec x,\tau) d\tau
  \right].
}

The expectation value of Polyakov loop $\langle L(\vec x) \rangle$ 
is an order parameter for confinement-deconfinement transition in the pure 
glue theory.\ Here,\ $P$ denotes path ordering of the exponential,\  
$g$ is the gauge coupling,\ and $\beta_{\tau} = 1/T$ denotes the extent of 
Euclidean time.\ $A_\mu(\vec x,\tau) = A^a_\mu(\vec x,\tau) \lambda^a$ is 
the vector potential.\ The $\lambda^a$  are the $N^2 - 1$ Hermitian 
generators of the $SU(N)$ algebra in the fundamental representation.\ 
$A_0(\vec x,\tau)$ is the time component of the vector potential at 
spatial position $\vec x$ and Euclidean time $\tau$.\ The gauge fields 
$A_\mu(\vec x,\tau)$ obey periodic boundary conditions in the Euclidean 
time direction,\ $A_\mu (\vec x,\beta_{\tau}) = A_\mu (\vec x, 0)$.\ These 
boundary conditions are maintained by a group of non-trivial gauge 
transformations~\cite{'tHooft:1977hy} that are periodic up to a constant twist 
matrix,\ $z \in SU(N)$,
\eq{
  \label{zntn}
  g(\vec x,\beta_{\tau}) = z\, g(\vec x,0).   
}

These  matrices $z$ form the center $Z(N)$ of the gauge group $SU(N)$,\ 
where $Z(N)$ is a cyclic group of order $N$.\ Thus the pure $SU(N)$ gauge 
theory at finite temperature has the complete symmetry
${\cal G}\times Z(N)$,\ where ${\cal G}$ is the group of strictly periodic 
gauge transformations. 

However,\ the Polyakov loop (which characterises the phases of pure gauge theory) 
transforms non-trivially under the $Z(N)$ transformations,\ though the Euclidean action 
is invariant under the transformations.\ Under the global $Z(N)$ symmetry transformations,\  
$L(\vec x)$ transforms as
\eq{
  L(\vec x) \rightarrow z\, L(\vec x),
}
where,
\eq{
  z = \exp(2\pi i n/N) {\mathbbm{1}}~\in Z(N),~~ n \in 
  \left\{0,1,2,\cdots,N-1 \right\}.
}

For temperatures above the critical temperature $T_c$,\ the high temperature 
phase or the deconfining phase is characterised by $\langle  L(\vec x) \rangle 
\neq 0$,\ corresponding to the finite free energy of an isolated heavy test 
quark,\ and thus breaks the $Z(N)$ symmetry spontaneously.\ At temperatures 
below $T_c$ (in the confining phase),\ $\langle  L(\vec x) \rangle  = 0$, 
thereby restoring the $Z(N)$ symmetry~\cite{McLerran:1981pb}.

We now come to the effect of matter fields in fundamental representation 
on $Z(N)$ symmetry. In the fundamental representation of $SU(N)$,\ the quark 
fields transform as,
\eq{
  \Psi \rightarrow g\, \Psi.
}
As a result the fermion part of the action is not invariant under the twisted transformations 
pertaining to the $Z(N)$ symmetry.\ Thus in full QCD,\ the fermion part of the action breaks 
the $Z(N)$ symmetry while the pure gauge part respects it.

The effect of quarks on the $Z(N)$ symmetry has been discussed in detail 
in~\cite{Belyaev:1991cw,KorthalsAltes:1994hq}.\ It has been advocated that one can take 
the effect of quarks in terms of explicit breaking of 
$Z(3)$ symmetry~\cite{KorthalsAltes:1994hq,Pisarski:2000eq,Dumitru:2000in}.\ This leads 
to a unique ground state and two (possible) meta-stable states in the deconfinement
phase~\cite{Dixit:1991et}.\ It has also been argued that the effects of quarks in terms of 
explicit symmetry breaking may be small,\ and the pure glue theory may be a good 
approximation~\cite{Dumitru:2000in}.\ However,\ we will see later that the explicit 
symmetry breaking due to quarks is too large for the meta-stable states to exist in the 
neighbourhood of $T_c$.


\section{Numerical Parameters and Studies}

In principle meta-stable states should ocuur in any simulation run. Given enough
time the system will explore all of the configuration space. But the probability 
that an arbitrary initial configuration temporarily "thermalising" to a meta-stable 
state is very small. Once the system thermalises to the absolute ground state the 
probability that it fluctuates to the meta-stable state decreases exponentialy with the 
system size. So it is practical to choose initial configurations carefully so
that they thermalise to the meta-stable states before decaying to the ground state.
\ In order to have the system thermalized to a meta-stable state 
(${\mathrm{Re}}\, L < 0$ and $|{\mathrm{Im}}\, L| > 0$),\ the trial configuration 
should not be far away from this state.\ Otherwise,\ it will thermalize to the absolute 
ground state.\ Since a random configuration (termed as ``fresh'' in {\tt MILC} code) results 
in a stable state (${\mathrm{Re}}\, L > 0$),\ we need to select the initial configuration 
appropriately.\ As the meta-stable states arise from the $Z(3)$ symmetry so they are 
expected to be close to the $Z(3)$ rotation of the absolute ground state.\ This fact can be
used to find an initial configuration which may thermalize to the meta-stable 
state.\ To generate the gauge field configurations,\ we have used the {\tt MILC} code which 
uses the standard {\tt Hybrid R algorithm}~\cite{Gottlieb:1987mq}.\ We have used the 
similar simulation parameters as in~\cite{Machtey:2009wu,Karsch:2000ps}. 

As a first step,\ we have performed a pure gauge calculation on a lattice of 
size $16^3 \times 4$ near critical temperature 
($\beta = \beta_c = 5.6925$~\cite{Boyd:1995zg}) starting with a ``fresh lattice''.\ 
We then have selected one configuration each for the the following cases\---\
\begin{itemize}
\item[(i)]
${\mathrm{Re}}\, L \ll 0$ and ${\mathrm{Im}}\, L \gg 0$,\ i.e.\ $\theta \simeq 2\pi/3$.
\item[(ii)]
${\mathrm{Re}}\, L \ll 0$ and ${\mathrm{Im}}\, L \ll 0$,\ i.e.\ $\theta \simeq - 2\pi/3$.
\item[(iii)]
${\mathrm{Re}}\, L \ll 0$ and ${\mathrm{Im}}\, L \sim 0$,\ i.e.\ $\theta \simeq \pi$.
\item[(iv)]
${\mathrm{Re}}\, L \gg 0$ and ${\mathrm{Im}}\, L \sim 0$,\ i.e.\ $\theta = 0$.
\end{itemize}
The case (iii) has been taken into consideration additionally to study whether a 
meta-stable state could exist at $\theta = \pi$~\cite{Machtey:2009wu},\ and the 
case (iv) to check whether we obtain the similar stable state if we start with a 
``fresh lattice''.\ Another method,\ we have employed,\ is by using an initial 
configuration with all the temporal links on a fixed time slice set to 
$e^{\displaystyle \pm{2\pi i/3}}{\bf I}$.\ The pure gauge calculations have been 
performed using the {\tt MILC} code.\ We have considered  4 over-relaxation steps for 
each of the heat-bath iteration for updating a gauge configuration.

We have used each of the gauge configurations (one for each $\theta$) so 
obtained as an initial configuration to thermalize,\ and calculate the Polyakov 
loop in presence of 2 and 3-flavor quarks for a series of $\beta$ values,\ 
$5.2 \leq \beta \leq 6.0$.\ For this purpose,\ we have again used the {\tt MILC} code with 
dynamical staggered fermion action.\ The numerical calculations have been performed 
with quark mass $am_{u,d} = 0.01$,\ so that $m_{u,d}/T = 0.04$.\ Our micro-canonical 
time step size has been $\Delta\tau \sim 0.01$,\ and the trajectory length 
(micro-canonical time step $\times$ steps per trajectory) has been $\tau \sim 0.8$.\ 
Each gauge configuration has been analysed after 10 heat-bath iterations.\ For each 
$\beta$,\ we have collected 2500 gauge configurations.
%

%

\section{Results and Discussions}

\subsection{How to observe the meta-stable states?}

In this study the Polyakov loop expectation value $\left< L \right>$ has been
used to charecterise different possible states of the system. In the absolute
ground state $\left< L \right>$ is real and positive. In the meta-stable states 
$\left< L\right>$ is complex, with phase of $\left <
L\right>$ close to $\displaystyle\pm \frac{2\pi}{3}$. 

In a typical Monte-Carlo simulation,\ a sequence of statistically significant
configurations is generated.\ This sequence constitutes the Monte-Carlo history.\ 
In order to reduce the auto-correlations between the consecutive configurations,\ 
measurements are perfomed on every 10th configuration in the Monte-Carlo history. 
In all the different simulations,\ the initial configuration is thermalised in 
a few hundred Monte-Carlo steps.\ During the thermalisation,\ all the measured
observables change almost monotonically.\ Once the system is thermalised,\ all the 
physical observables fluctuate around their respective average values.\ We compute 
the histogram (probability distribution) of the Polyakov loop,\ $P(L)$.\ The 
meta-stable states are local minima in the free energy which are stable against small
fluctuations.\ As a result,\ they appear as well defined peaks in $P(L)$.\ Distinct Multiple 
peaks in $P(L)$ correspond to different possible states of the system. 

\begin{figure}[h]
  \centering
  \subfigure[]
  {\rotatebox{270}{\includegraphics[width=0.34\hsize]{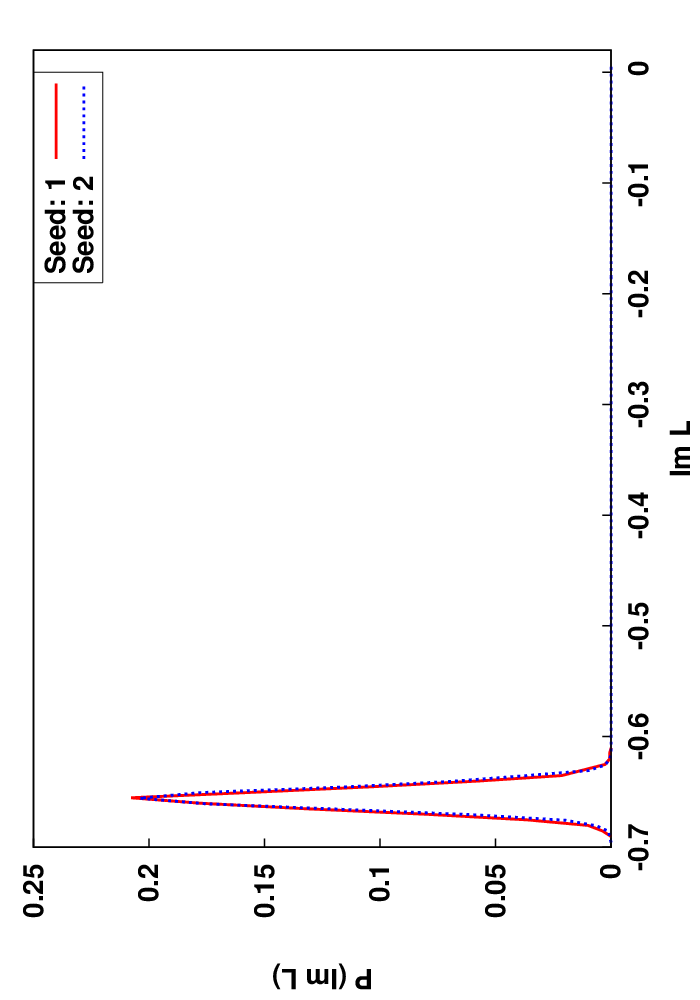}}
    \label{histo_Im_L_1a}
  }
  \subfigure[]
  {\rotatebox{270}{\includegraphics[width=0.34\hsize]{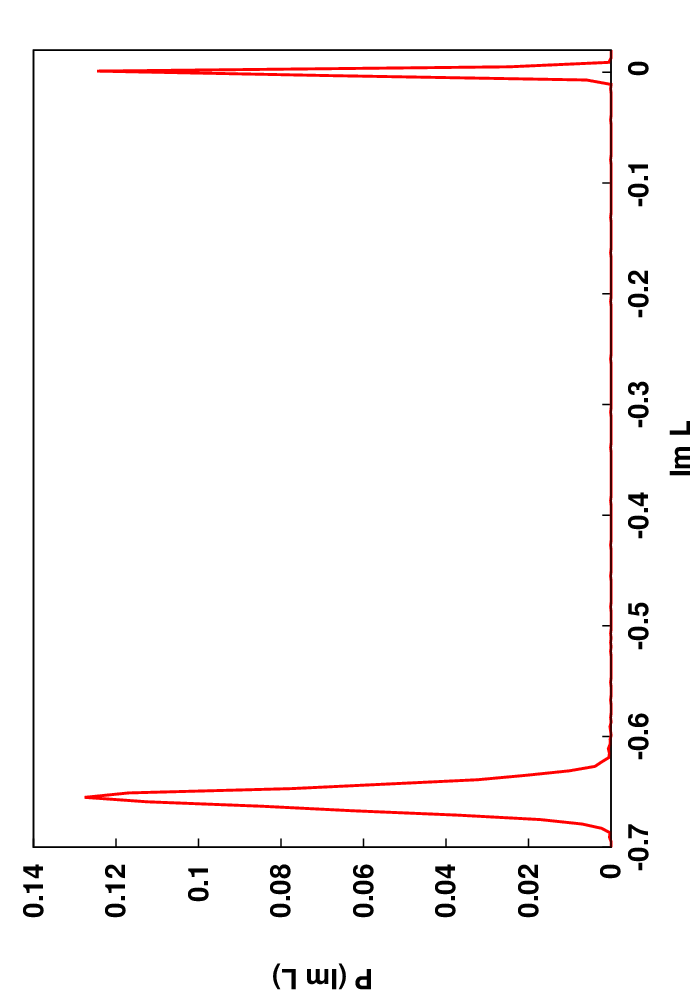}}
    \label{histo_Im_L_1b}
  }
  \caption{Histograms of ${\mathrm{Im}}\, L$ vs. no. of trajectories 
    (a) Single peak at ${\mathrm{Im}}\, L \neq 0$ indicating a meta 
    stable state only,\ 
    (b) Double peak at ${\mathrm{Im}}\, L \neq 0$ and ${\mathrm{Im}}\, L = 0$ 
    indicationg a meta stable state decaying into the ground state.}
    \label{histo_Im_L_1}
\end{figure}

In Fig.~\ref{histo_Im_L_1a},\ we show the histogram of 
${\mathrm{Im}}\, L$ (imaginary part of $L$) for two simulations with different random 
number sequences.\ The two $P({\mathrm{Im}}\, L)$ distributions are similar given the 
small length (a few thousands) of the Monte-Carlo histories.\ The peak positions are at a 
non-zero value of ${\mathrm{Im}}\, L\sim - 0.65$ which imply a meta-stable state.\ This 
being a meta-stable state,\ it will decay to the ground state if the simulations were 
continued further.\ In Fig.~\ref{histo_Im_L_1a}),\  we see only one peak because the state 
did not decay during our run time.\ With another sequence of random numbers,\ the 
meta-stable decays to the ground state.\ This results in a second peak in 
$P({\mathrm{Im}}\, L)$ at ${\mathrm{Im}}\, L\sim 0$ as seen in 
Fig.~\ref{histo_Im_L_1b}.\ Note that the peak position of the meta-stable state is 
at ${\mathrm{Im}}\, L\sim - 0.65$ which is same as that in Fig.~\ref{histo_Im_L_1a}.\ We 
find that $P({\mathrm{Re}}\, L, {\mathrm{Im}}\, L)$ is independent of thermalisation
apart from an overall normalisation factor. Random processes such as thermalisation can not
give rise to such peaks. 

Ideally the system should be allowed to explore all possible states in a single run.\ If there 
are meta-stable states,\ the distribution $P({\mathrm{Re}}\, L, {\mathrm{Im}}\, L)$ will 
have three distinct peaks.\ The ratio of the peak heights of the meta-stable state to the 
ground state will be given approximately by the exponential of the difference in free 
energies of these two states.\ Thereby, this ratio will be vanishingly small for large 
volumes.\ This translates to the vanishing probability of observing these meta-stable states 
starting from any arbitrary initial configuration.\ These states can only be observed by 
starting from suitable initial configurations.\ We believe such a method of study has no 
bearing on the properties of the meta-stable states.\ This is evident as we observe almost 
identical distributions $P({\mathrm{Re}}\, L, {\mathrm{Im}}\, L)$ for different choices of 
initial configurations and random number sequences as they all thermalise to a particular 
meta-stable state.

We have already mention that the physics we study here is different from the meta-stability 
observed near the critical point of a first order transition.\ For a first order 
transition,\ meta-stable (stable) state becomes stable (meta-stable) as the system passes
through the critical point.\ The meta-stable states are observed close to the
transition point.\ On the other hand,\ the $Z(3)$ meta-stable states exist 
away from the phase transition point as we discuss below. 

\subsection{Studies with a given volume}

\begin{figure}[h]
  \centering
  \subfigure[]
  {\rotatebox{270}{\includegraphics[width=0.34\hsize]{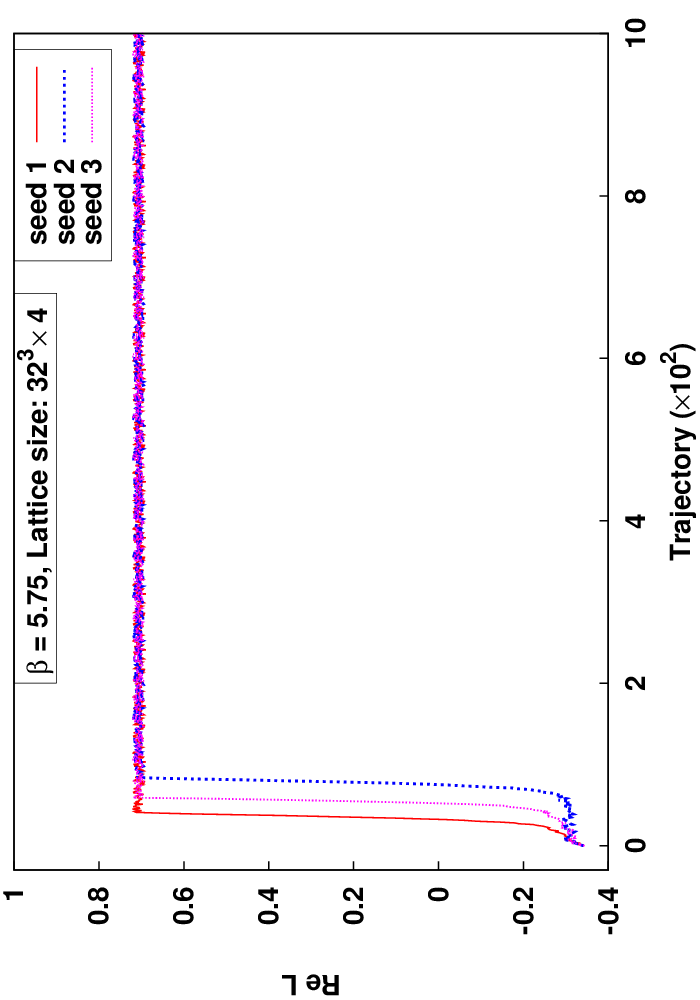}}
    \label{fig_beta_5.75_32x4_re_L_vs_trajec_no_seed_03}
  }
  \subfigure []
  {\rotatebox{270}{\includegraphics[width=0.34\hsize]{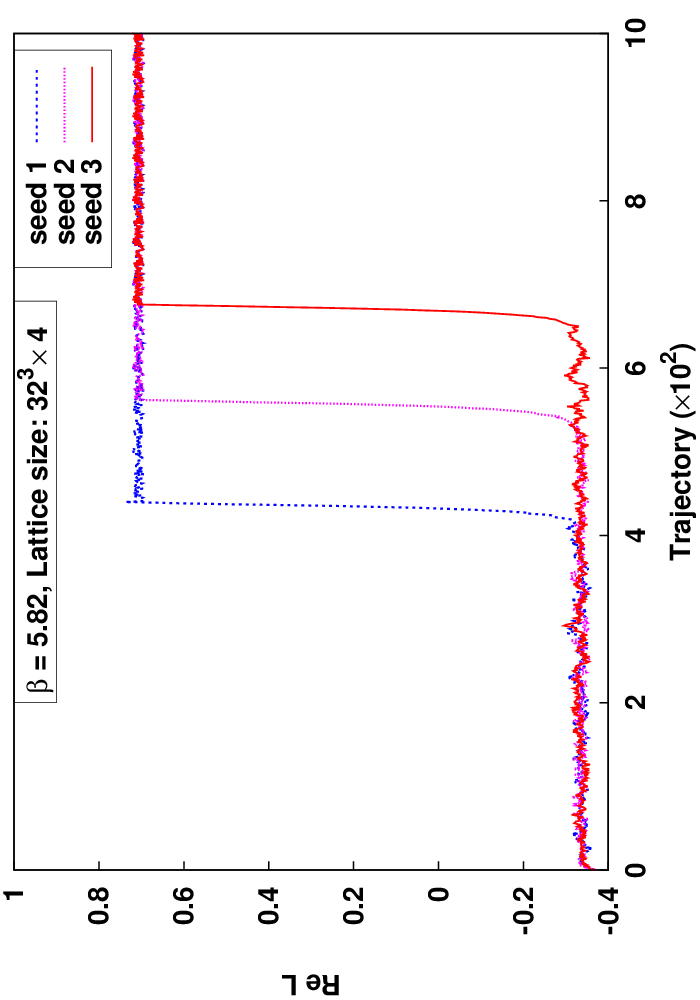}}
    \label{fig_beta_5.81_32x4_re_L_vs_trajec_no_seed_03}
  }
    \caption{${\mathrm{Re}}\, L$ vs. number of trajectories for three different 
      randomly chosen seeds at
      (a) $\beta = 5.75$, and 
      (b) $\beta = 5.82$ 
      for 2-flavor case.}
    \label{fig_re_L_vs_trajec_no_seed_03}
\end{figure}

In our simulations for $N_s = 32, N_\tau = 4$, with $\beta$ values up to 
$\beta_m \sim 5.80$,\ the system thermalises to the ground state irrespective of 
the initial configurations (listed in the previous section)  
(Fig.~\ref{fig_beta_5.75_32x4_re_L_vs_trajec_no_seed_03}).\ The system thermalises 
to the meta-stable states only for $\beta \ge \beta_m$ 
(Fig.~\ref{fig_beta_5.81_32x4_re_L_vs_trajec_no_seed_03}).\ One can argue 
that the meta-stable states are not observed below $\beta_m$ because the thermal 
flucutations are large compared to the barrier height between these states and 
the stable state.\ In effect this situation can be described by an effective potential without 
any local minima.\ We mention here that the chiral condensate takes slightly higher 
value in the meta-stable states than that of the ground state.

\begin{figure}[]
  \centering
      {\rotatebox{270}{\includegraphics[width=0.45\hsize]{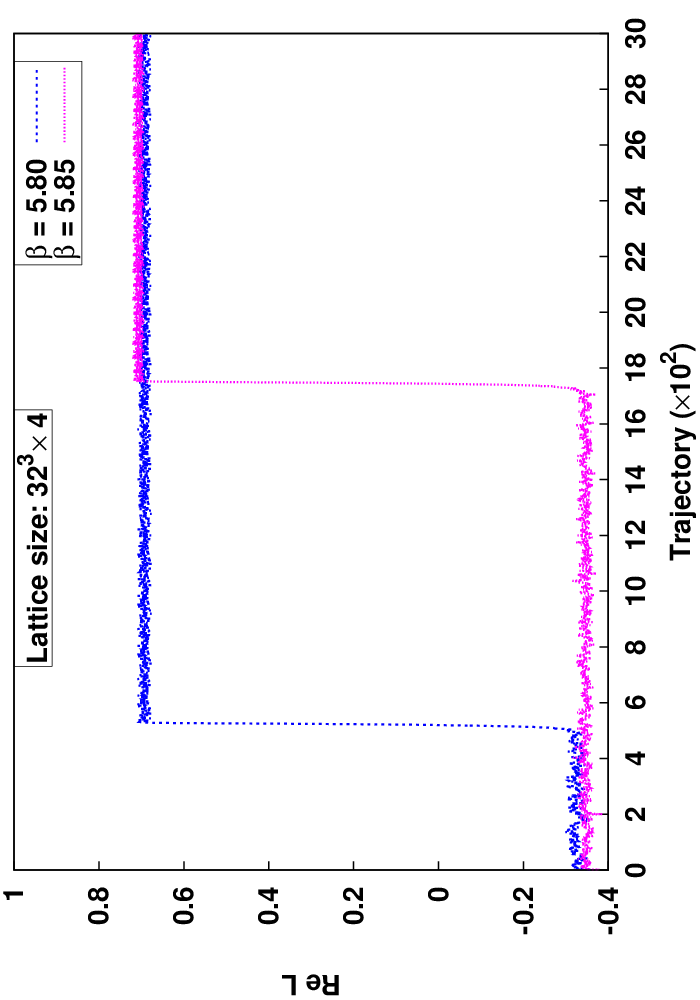}}
      }
      \caption{${\mathrm{Re}}\, L$ vs. number of trajectories at $\beta = 5.80$,\ 
        and $\beta = 5.85$ on a $32^3 \times 4$ lattice.}
      \label{fig_beta_580_and_585_32_cube_by_4}
\end{figure}

\begin{figure}[]
  \centering
  \subfigure[]
  {\rotatebox{270}{\includegraphics[width=0.30\hsize]{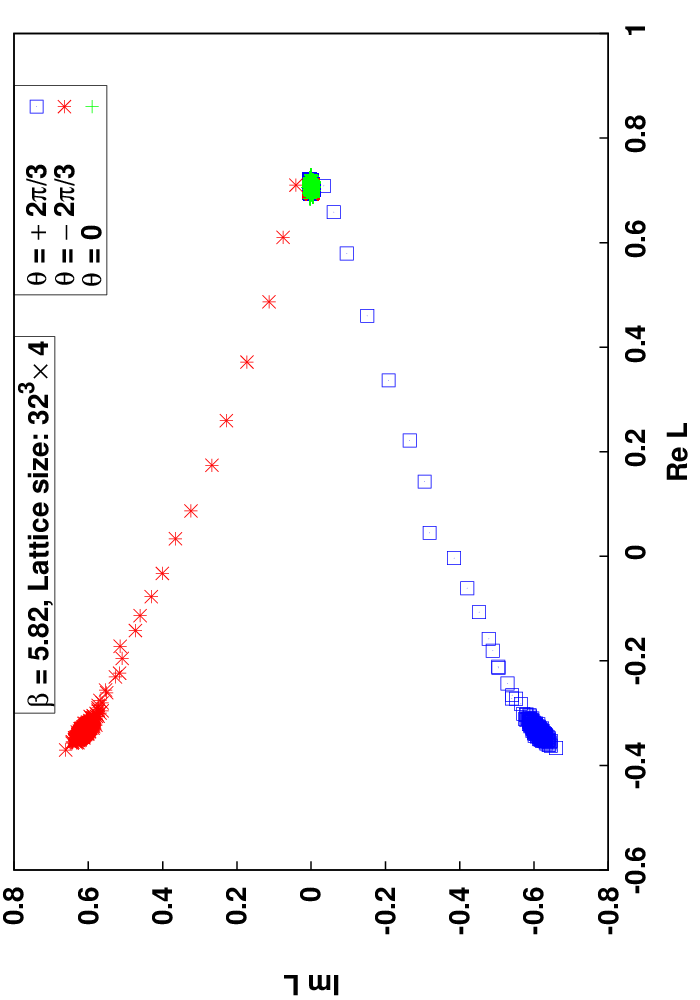}}
	\label{fig_ploop_w_var_theta_a}
	}
  \subfigure []
  {\rotatebox{270}{\includegraphics[width=0.30\hsize]{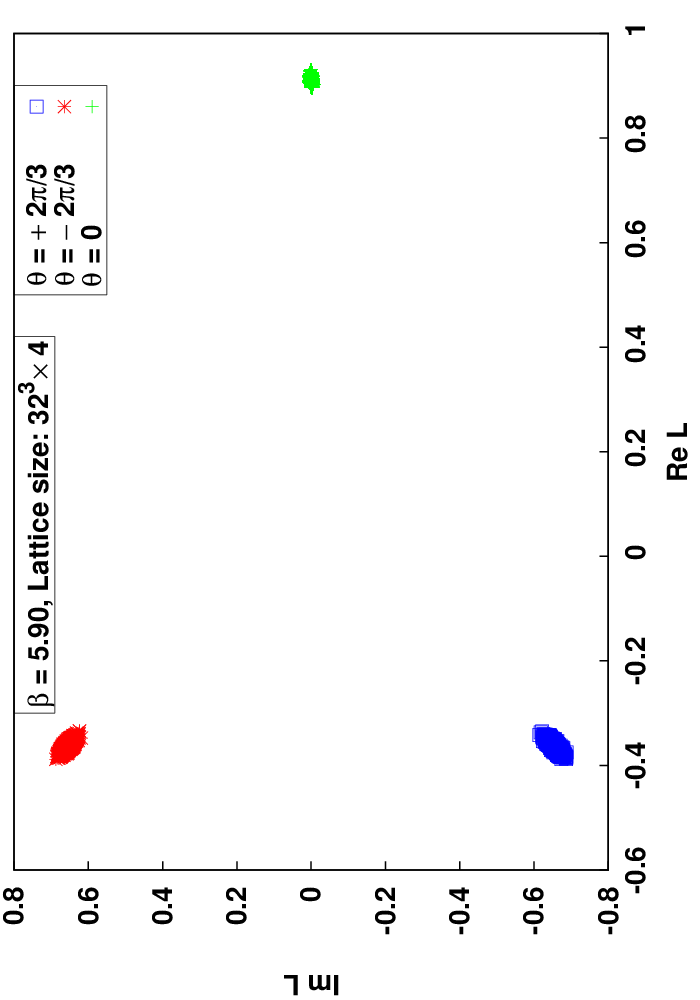}}
	\label{fig_ploop_w_var_theta_b}}
  \caption{Imaginary vs. real part part of Polyakov loop for two different 
    values of $\beta$.}
  \label{fig_ploop_w_var_theta}
\end{figure}
%


Higher $\beta$ corresponds to higher temperature and larger quark 
mass~\cite{Gottlieb:1995,Karsch:2000ps}.\ In the pure $SU(3)$ gauge theory,\ 
the barrier height between the $Z(3)$ states increases with temperature.\ This
should also be true in full QCD,\ since gluons dominate at higher temperatures.\ So 
it is expected that the barrier height between the meta-stable state and the absolute 
ground state increases with $\beta$.\ As a result,\ the system in meta-stable state should 
spend on an average longer "Monte-Carlo time" in the meta-stable state.\ Considering 
different initial configurations and random number sequences (for a particular $\beta$ and 
$m_q$),\ we calculate the average Monte-Carlo time the system spends in the meta-stable 
states.\ We find that this time monotonically increases with $\beta$.\ For example,\ in 
Fig.~\ref{fig_beta_580_and_585_32_cube_by_4} we have shown the case for 
$\theta = - 2\pi/3$,\ and $\beta = 5.80$ and $5.85$.\ For $\beta = 5.85$,\ the system
stays longer in the meta-stable state than $\beta = 5.80$.\ For $\beta=5.90$,\ the meta-stable 
state did not decay during the simulation we considered.\ In 
Figs.~\ref{fig_ploop_w_var_theta},\ we show 
${\mathrm{Re}}\, L$ vs. ${\mathrm{Im}}\, L$ for different and $\beta$ values.\ We see 
that for $\beta = 5.90$ (Fig.~\ref{fig_ploop_w_var_theta_b}),\ there are no fluctuations 
between different minima.\ With various different trial configurations and different seeds,\ we 
found only two meta-stable states.\ No other (meta)stable state was observed in our 
simulations,\ such as $\theta = \pi$~\cite{Machtey:2009wu}.

\subsection{Volume and $N_\tau$ studies}

\begin{figure}[]
  \centering
    \subfigure[]
      {\rotatebox{270}{\includegraphics[width=0.34\hsize]{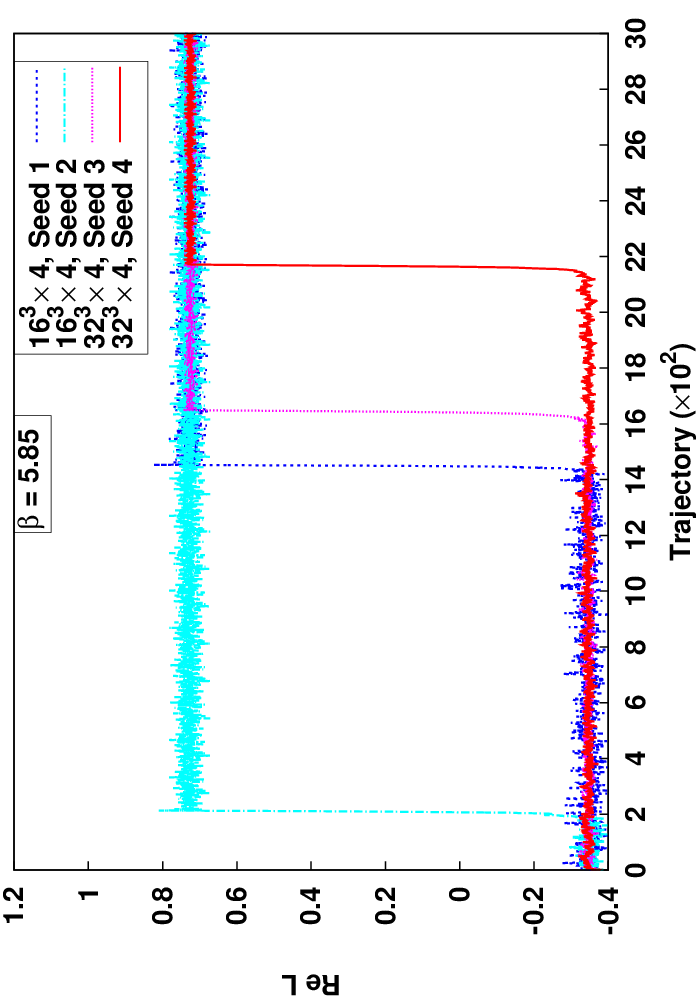}}
      \label{fig_beta_592_16_and_32_cube_by_6_a}
	}
      \subfigure[]
      {\rotatebox{270}{\includegraphics[width=0.34\hsize]{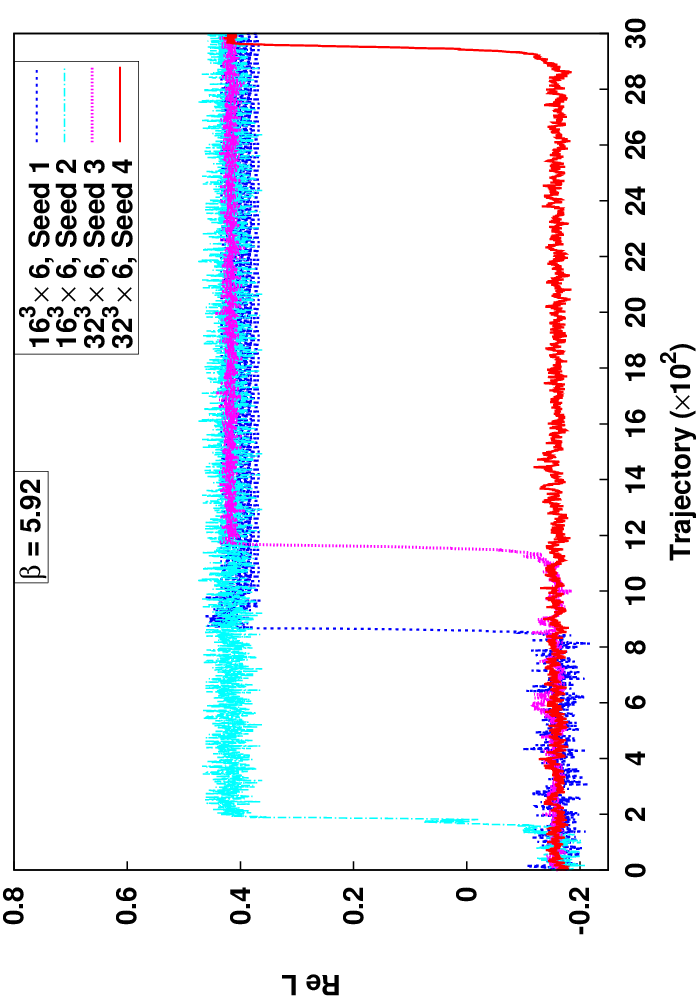}}
      \label{fig_beta_592_16_and_32_cube_by_6_b}
	}
      \caption{${\mathrm{Re}}\, L$ vs. number of trajectories for 
        $V=16^3,\ 32^3$ at (a) $\beta = 5.85$, $N_\tau=4$ and 
        (b) $\beta=5.92$, $N_\tau=6$.}
      \label{fig_beta_592_16_and_32_cube_by_6}
\end{figure}

We extend our studies to different spatial voulmes,\ $V$'s,\ and temporal extent,\ $N_\tau$'s,\ 
in order to investigate whether the finite volume effects play any significant role in the 
observation of the meta-stable states.\ We conside $V = 16^3, 24^3, 32^3$ and 
$N_\tau = 4, 6$.\ For a particular choice of ($V$, $N_\tau$, $n_f$ and  $m_q$),\ we always 
find a $\beta_m$ above which there are meta-stable states.\ For a fixed $N_\tau$.\ the value 
of $\beta_m$ remains almost the same as volume changed.\ The average value of $|L|$ in 
the meta-stable(stable) state remains the same as we increase $V$ from $16^3 \--\ 32^3$ 
as seen in Figs.~\ref{fig_beta_592_16_and_32_cube_by_6}.\ For the different volumes we 
considered,\ the Polyakov loop susceptibility does not have any dependence on the size of 
the system.\ This is expected since at $\beta=\beta_m$,\ the system is away from the 
transition region.\ The corresponding correlation length is expected to be very small compared 
to the system size.\ So we hardly see any finite size effects.\ In 
Fig.~\ref{fig_beta_592_16_and_32_cube_by_6},\ it can be seen that the average value of $|L|$ 
and it's fluctuations decrease when $N_\tau$ changes from $4$ to $6$.\ This behavior is observed 
in both the meta-stable and the ground states.\ In our estimate for $\beta_m$,\ we find the value 
increases with $N_\tau$.\ For example, for $V=32^3$ we find that $\beta_m \sim 5.80 (5.89)$ 
for $N_\tau = 4 (6)$.\ The $\beta-$function~\cite{Gottlieb:1995} results show that a 
fixed temperature scale corresponds to a higher $\beta$ for larger $N_\tau$ in the
$\beta$ range of our study.\ Interestingly,\ we find that the two $\beta_m$'s obtained for 
$N_\tau = 4, 6$ correspond to the temperature scales close to each other.\ However,\ a 
quantitative comparison between different $N_\tau$ results will require higher statistics than 
the present simulations.

In all the volumes we study,\ we find a similar pattern in the behavior of the Polyakov 
loop in the meta-stable and stable states.\ As expected,\ $\beta_m$ increases with 
$N_\tau$.\ But we would like to mention that our study of thermodynamic and continuum 
limit is far from complete.\ With the increase in the size of the system,\ the meta-stable 
states should take longer {\em average} Monte-Carlo time to decay to the stable state.\ 
In our simulations with a small number of seeds,\ we do see that the meta-stable states 
take longer average MC time to decay with the increase in volume.\ However, a quantitative
estimate of the volume dependence of the average decay time will need a large number of 
simulations with different seeds.\ We mention here that same $\beta$ corresponds to lower 
physical temperature for higher $N_\tau$.\ This can lead to smaller decay time for the 
meta-stable states for higher $N_\tau$ (see Fig.~\ref{fig_beta_592_16_and_32_cube_by_6}).

\subsection{$n_f$ and $m_q$ dependence}

\begin{figure}[h]
  \centering
  {\rotatebox{270}{\includegraphics[width=0.60\hsize]{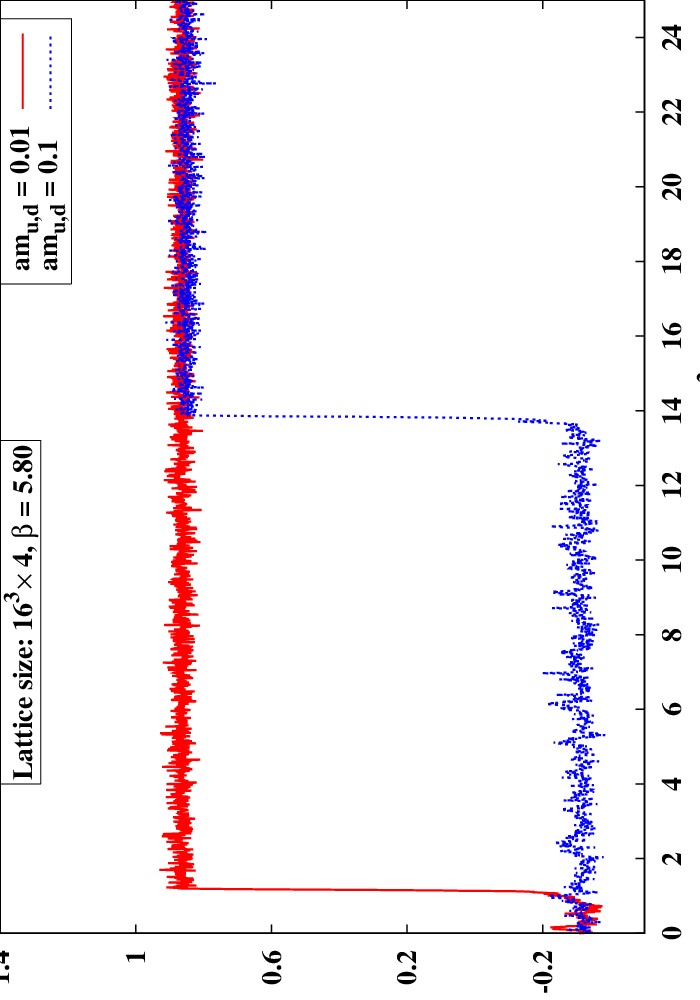}}}
  \caption{${\mathrm{Re}}\, L$ vs. no. trajectories at $\beta = 5.80$ with masses 
    at $am_{u,d} = 0.01$ and $am_{u,d} = 0.1$.}
  \label{fig_ploop_w_low_mass}
\end{figure}

We have repeated our simulations for $3-$flavors of degenerate quarks.\ In this
case,\ the meta-stable states appear at higher values of $\beta_m$ compared to that 
of $2-$flavor.\ Apart from higher $\beta_m$,\ all other findings are similar to 
those of 2-flavor case.\ So,\ we have not shown any results for $3-$flavor here.\ 
We have also considered higher values of $(m/T)$ in our simulations.\ For infinitely 
heavy quark masses,\ there is no explicit symmetry breaking of $Z(3)$.\ So,\ we expect 
that higher the quark mass smaller is the explicit symmetry breaking.\ For a larger 
mass $am_{u,d} = 0.1$,\ the Monte Carlo histories for $\beta = 5.80$ has been shown 
in Fig.~\ref{fig_ploop_w_low_mass} together with $am_{u,d} = 0.01$ for the same 
$\beta$ and seed.\ We see that the meta-stable state with $am_{u,d} = 0.1$ decays 
later than that of $am_{u,d} = 0.01$.\ This suggests that $\beta_m$ is smaller for 
higher $(m/T)$ and approaches the pure gauge transition point for $m_q \rightarrow
\infty$.

$(m/T)$ is fixed in our simulations.\ But for realistic simulations,\ one should fix 
the quark masses.\ This amounts to considering a smaller $(m/T)$ at higher $\beta$.\ 
Smaller $(m/T)$ will decrease the barrier height between the meta-stable and ground 
state.\ On the other hand,\ the pure gauge effects will make the barrier height
grow at higher $\beta$.\ We expect that the pure gauge effects will take over the
explicit symmetry breaking term for $\beta > \beta_m$.\ The value of $\beta_m$ we
get,\ should be the lower bound for all the quark masses for which 
$am_{u,d}(\beta_m) \le 0.01$.

The quark masses we have considered are larger compared to that of the physical $u,d$ 
quarks.\ So,\ if we ignore the strange quark,\ the $\beta_m$ for QCD with $u,d$ 
quarks will be higher than the value we got from our 2-flavor simulations.\ Inclusion 
of strange quark will only increase the $\beta_m$.\ From this,\ we can conclude that 
$\beta_m$ for realistic case of light $u, d$ quarks,\ and a heavy strange quark will 
be larger than the value for our $2-$flavor calculations.\ Note that $\beta_m$ is 
larger than the $\beta_c$ for the Quark-Hadron transition.\ As a result
the meta-stable states do not appear near the Quark-Hadron transition but at larger
temperatures.

\subsection{Estimate of temperature scale $T_m$ for the meta-stable states}

Now we make an estimate of the temperature $T_m$ corresponding to $\beta_m$ and 
$am_q$.\ This requires the $\beta-$function which relates these bare parameters to the 
lattice cutoff,\ $a$.\ The $\beta-$function for $N_f  = 2$ which is appropriate for the 
action we have used has been studied in detail in~\cite{Gottlieb:1995} in the range 
$5.2 \le \beta \le 5.6$.\ However,\ the values of $\beta_m$ which is of interest to us 
lie outside this range.\ But we naively used the same $\beta-$function as 
in~\cite{Gottlieb:1995} to make an estimate of $T_m$.\ This gives us 
$T_m = T (\beta_m = 5.80, am_q = 0.01) \simeq 750$~MeV.\ This temperature should be 
compared with the phase transition temperature which is about $~200$~MeV.\ For a better 
estimate,\ we should have lattice results for the $\beta(a)$ close to $\beta_m$.\ However,\ 
we believe that this will not drastically reduce $T_m$.\ As we have argued before,\ 
$\beta_m$ will be larger for realistic quark masses with $N_f = 2+1, 3$.\ So,\ it is 
likely that the actual value of $T_m$ will be larger than our estimate.

Previous studies have suggested that the explicit symmetry breaking due to quarks is 
small~\cite{Dumitru:2000in}.\ But as we can see here that the explicit symmetry 
breaking is strong enough to make the $Z(3)$ states with non-zero phase angle unstable 
below $T_m$.\ The value of $T_m > 750$~MeV suggests that this states will not be excited 
in RHIC.\ There is a possibility,\ however,\ that they will be observed at LHC.\ During 
thermalization a local region or the whole of the fireball may be trapped in one of the 
meta-stable states.\ The regions in meta-stable phase can then decay through bubble 
nucleation or via spinodal decomposition depending on dynamical evolution of the 
system~\cite{Dixit:1991et,Ignatius:1991nk}.
It would be interesting to explore the possible consequence of the decay of the
meta-stable phases. For temperature above $T_m$ the system is dominated by gluons
and $Z(3)$ is effectively restored. This allows for various types of domain walls,
static or otherwise~\cite{us}. These solutions become unstable below $T_m$.


\section{conclusion}

We have done full QCD simulations with $N_f=2, 3-$flavor of degenerate quarks to study
meta-stable states in the neighborhood of $T_c$.\ It has been found that the
$\beta$ value above which the meta-stable states appear is close to critical value of
$\beta$ for the pure gauge confinement-deconfinement transition.\ We estimate
the temperature scale to be $T_m > 750 $~MeV (for $N_f=2$) above which meta-stable
states can appear.\ So,\ we expect that these meta-stable states may be observed at LHC.\
There is a non-trivial change in the shape of the effective potential for
the Polyakov loop as the system cools through $T_m$.\ Below $T_m$,\ there are no local 
minima in the effective potential.\ So,\ the meta-stable states, $Z_3$ domain walls become 
unstable.\ This may have important consequences for systems such as the fireball in 
heavy-ion collisions or the early Universe.


\acknowledgements

We would like to thank A. M. Srivastava and Saumen Datta for useful discussions and 
comments. All our numerical computation has been performed at Annapurna (SGI Altix), a 
supercomputer based at the Institute of Mathematical Sciences, Chennai, India.\ This work is 
based on the {\tt MILC} collaboration's public lattice gauge theory code 
(version 6)~\cite{milc}.



\vskip 25pt

\phantomsection
\addcontentsline{toc}{chapter}{References}

\centerline{\bf REFERENCES} \vskip -15pt

\end{document}